# Non-Statistical Effects in Neutron Capture

P. E. Koehler[a], T. A. Bredeweg[b], K. H. Guber[a], J. A. Harvey[a], J. M. O'Donnell[b], R. Reifarth[c], R. S. Rundberg[b], J. L. Ullmann[b], D. J. Vieira[b], D. Wiarda[a], and J. M. Wouters[b]

[a]*Oak Ridge National Laboratory, Oak Ridge, TN, USA.*
[b]*Los Alamos National Laboratory, Los Alamos, NM, USA*
[c]*GSI, Planckstr. 1, 64291 Darmstadt, Germany*

**Abstract.** There have been many reports of non-statistical effects in neutron-capture measurements. However, reports of deviations of reduced-neutron-width ($\Gamma_n^0$) distributions from the expected Porter-Thomas (PT) shape largely have been ignored. Most of these deviations have been reported for odd-A nuclides. Because reliable spin ($J$) assignments have been absent for most resonances for such nuclides, it is possible that reported deviations from PT might be due to incorrect $J$ assignments. We recently developed a new method for measuring spins of neutron resonances by using the DANCE detector at the Los Alamos Neutron Science Center (LANSCE). Measurements made with a $^{147}$Sm sample allowed us to determine spins of almost all known resonances below 1 keV. Furthermore, analysis of these data revealed that the $\Gamma_n^0$ distribution was in good agreement with PT for resonances below 350 eV, but in disagreement with PT for resonances between 350 and 700 eV. Our previous $(n,\alpha)$ measurements had revealed that the $\alpha$ strength function also changes abruptly at this energy. There currently is no known explanation for these two non-statistical effects. Recently, we have developed another new method for determining the spins of neutron resonances. To implement this technique required a small change (to record pulse-height information for coincidence events) to a much simpler apparatus: A pair of $C_6D_6$ γ-ray detectors which we have employed for many years to measure neutron-capture cross sections at the Oak Ridge Electron Linear Accelerator (ORELA). Measurements with a $^{95}$Mo sample revealed that not only does the method work very well for determining spins, but it also makes possible parity assignments. Taken together, these new techniques at LANSCE and ORELA could be very useful for further elucidation of non-statistical effects.

## INTRODUCTION

Neutron capture and scattering on intermediate to heavy nuclides at low energies are expected to proceed through the formation and decay of highly complicated compound nuclear states. Hence, these cross sections should be well described by the nuclear statistical model, and the reduced neutron width ($\Gamma_n^0$) and resonance spacing distributions should agree with the Porter-Thomas (PT)[1] and Wigner[2] distributions, respectively, as predicted by random matrix theory[3]. Deviations from the expected distributions often are referred to as non-statistical effects.

The expectation that $\Gamma_n^0$ values follow a PT distribution arises from the fundamental assumptions that the expansion coefficients of the compound nuclear wave function follow a Gaussian distribution with zero mean (i.e., that they are "statistical"), that these coefficients are real (because, due to time-reversal invariance, the reduced width amplitudes have been shown to be real), and that neutron scattering is a single-channel process at these energies (elastic scattering only). These assumptions result in the expectation that reduced neutron widths for resonances of a single $J^\pi$ will be distributed according to a $\chi^2$ distribution with one degree of freedom, or the PT distribution.

Tests of the PT distribution are hampered by several experimental challenges. For example, the PT distribution is dominated by small widths, which are the most difficult to observe in experiments. Also, the region of small widths is where deviations are expected to occur if they are caused by "new" physics.

In addition to the requirement that the data be complete (or a correction applied for missed small widths), it also is important that the data be pure. For example, a relatively small contamination of *p*-wave resonances in an *s*-wave distribution can skew the observed distribution towards smaller widths and make it appear that the data are in better agreement with PT than they actually are. Also, for odd-A targets there are two *s*-wave spins possible. Because it typically has not been possible to reliably separate resonances of the two spins, they have been combined while comparing the data to a PT distribution. However, implicit assumptions made while combining the data in this way may weaken the strength of tests of the PT distribution

Perhaps the most thorough test that $\Gamma_n^0$ values follow a PT distribution is that of Ref. 4 wherein data for *s*-wave resonances in nine even-even nuclides were examined. After a threshold was applied to exclude *p*-wave resonances as well as to attempt to ensure that no further correction for missed resonances was needed, the combined data were found to be in agreement with PT to within an uncertainty of 10%. On the basis of this and similar tests, and the strength of the assumptions underlying its validity, the PT distribution routinely is accepted as fact, and even often is used to correct the observed average level spacings and strength functions for missed resonances. However, there have been several reported cases[5-9] where the data deviate from a PT distribution. Unfortunately, all the cases of which we are aware are potentially problematic. Odd-A targets comprise all but one of the reported cases, and they have potentially more problems, as discussed above. The single even-even case ($^{232}$Th) suffers from the fact that it is near the peak of the *p*-wave strength function and so may suffer from impurity issues. For these reasons, better $J^\pi$ information for neutron resonances would be very helpful for checking reported non-statistical effects as well as searching for new ones.

We have developed two new techniques for measuring $J^\pi$ values of neutron resonances. We describe these methods in the next section. Following this, we describe the striking non-statistical effect observed in the first case in which we applied one of these techniques.

# TWO NEW TECHNIQUES FOR DETERMINING NEUTRON RESONANCE SPINS AND PARITIES

Both techniques are based on using information contained in the γ-ray cascade following neutron capture. Consider the case of $^{147}$Sm + n. Because the ground-state spin of $^{147}$Sm is $I^\pi = 7/2^-$, s-wave neutrons lead to $3^-$ and $4^-$ states in $^{148}$Sm. In a very simple model in which only dipole transitions can occur, at least three γ –ray transitions are required to reach the $0^+$ ground state from a $3^-$ excited state, whereas a minimum of four transitions are required in the case of a $4^-$ state. Hence, in this very simple model, $3^-$ resonances will have a minimum multiplicity of 3 and $4^-$ resonances a minimum multiplicity of 4. In reality, the abundance of available transitions and the existence of other multipolarities will both broaden the multiplicity distributions as well as decrease the difference between average multiplicities for $3^-$ and $4^-$ resonances[10,11].

A detector having near 100% efficiency that is composed of many individual elements can be used to measure average multiplicities and hence to determine resonance spins, as demonstrated in Ref. 12 in which spins of 91 $^{147}$Sm + n resonances below 900 eV were determined. Also, a much simpler apparatus composed of only two γ-ray detectors has been shown to be able to determine spins in experiments[10] using several odd-A targets. In Ref. 10, the same basic idea that the higher spin will have higher multiplicity is put to work: Resonances with the higher spin will result in more coincidences between the two detectors and, at the same time, a softer singles spectrum (because the same total energy is shared by more γ rays for the higher spin) in each detector. Therefore, the ratio of "hard" singles to all coincidences can be used to separate resonances of the two spins. We have invented new and improved versions of the former technique in an experiment at LANSCE and of the latter technique in an experiment at ORELA. Because both LANSCE and ORELA are white neutron sources, data across the entire range of energies were taken in single experiments, and the time-of-flight technique was used to determine the energy of the incident neutrons.

## The DANCE at LANSCE Experiment

The Detector for Advanced Neutron Capture Experiments (DANCE)[13,14] at the Manuel Lujan Jr. Neutron Scattering Center (MLNSC) at LANSCE was used to measure spins of neutron-capture resonances in $^{147}$Sm+n. Details of the experiment and results are given in Ref. 15. DANCE is a 4π BaF2 ball made from 160 individual detector elements. A 10.410-mg sample of metallic samarium, enriched to 97.93% in $^{147}$Sm, was placed in the beam line vacuum at the center of DANCE, 20 m from the neutron production target. Neutron capture events were recorded using separate transient digitizers for each of the 160 detectors. These events were sorted off line and an overall cut on the total γ-ray energy, $E_\gamma$ = 3–8 MeV, was used to restrict events to those in the range expected from $^{147}$Sm(n,γ) reactions. Then, average multiplicities were calculated over neutron energy ranges corresponding to the resonances. The resulting average multiplicities, in general, fell into two groups corresponding to the two s-wave resonance spins. This procedure mirrors the one used in Ref. 12 and resulted in good agreement with spins assigned in that reference for well resolved resonances. However, due to the relatively long proton pulse (FWHM~125 ns) and long tails from the neutron moderators at MLNSC, resonances become poorly resolved at rather low energies. As a result, above approximately 300 eV the average multiplicities tended towards values intermediate between those for the two spins for well resolved resonances, and it became very difficult to assign spins using this technique.

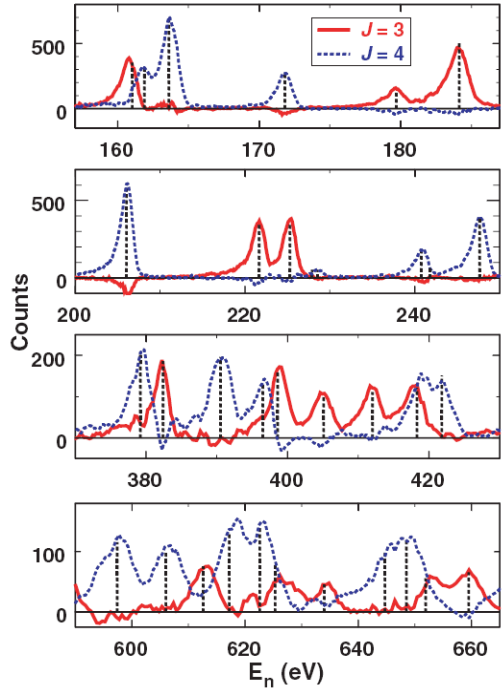

**FIGURE 1.** Two different linear combinations of multiplicities versus neutron energy from our DANCE data taken with a $^{147}$Sm sample. The solid curves were calculated using a combination which accentuates $J = 3$ resonances. Similarly, the dashed curves were calculated using a combination which accentuates $J = 4$ resonances. Dotted vertical lines indicate positions of resonances identified in previous work. The data have been smoothed over three to five channels to reduce statistical fluctuations. See Ref. 15 for details.

To overcome these difficulties, we invented a new technique based on using two different linear combinations of the measured multiplicities. The technique is explained in full in Ref. 15. In essence, it involves subtracting the prototypical multiplicity distribution for each spin from the data to generate two curves which peak at resonances corresponding one or the other of the two s-wave spins. Resonance spin values can be determined by simple inspection of these curves as shown in Fig. 1. This new technique works better than that of Ref. 12 for several reasons. First, it makes use more information: Both the shapes as well as the averages of the multiplicity distributions are used. Second, the two curves wax and wane with the inherent resonance structure of the data thereby effectively improving the resolution of resonances with different spins. In fact, using this technique we were able to identify six previously unknown "doublets". Third, it avoids division by the net total number of counts, as is needed for calculating average multiplicities. As a result of this division, average multiplicities calculated off resonance peaks can be very noisy because the net counts are small in these regions.

Overall, using this new technique we were able to make 41 new spin assignments for the 140 known resonances below $E_n$ = 1 keV. As a result, only 9 observed resonances below 700 eV remain without firm spin values.

## The CINDORELA Experiment

Measurements at ORELA were made with a pair of $C_6D_6$ γ-ray detectors on flight path 6 in the 40-m station. This new apparatus is basically a clone of the one on flight path 7 in the 40-m station with which we have measured numerous (*n*,γ) cross sections. Being a new apparatus, it was in need of a name, so we christened it the Capture of Incident Neutrons Detector at ORELA, or CINDORELA. One small change with respect to the old apparatus is that we now record pulse-height data for coincidence events. Our success measuring spins with the DANCE detector lead us to try the simple two-detector technique of Ref. 10, despite the fact that the coincidence rate with CINDORELA is quite low.

The sample for the initial experiment was 3.669 g of metallic molybdenum enriched to 96.47% in $^{95}$Mo. Separate sample-out background measurements were made and subtracted from the sample-in data. Separate two-dimensional spectra (time-of-flight, or neutron energy versus pulse height, or γ-ray energy) were constructed for both singles and coincidences by replaying the data off line.

Because $^{95}$Mo is near the peak of the *p*- and valley of the *s*-wave neutron strength functions, it is a much more difficult case than $^{147}$Sm (which is near the peak of the *s*-wave strength function and hence only *s*-wave resonances are seen at low energies). For $^{95}$Mo, there are six $J^{\pi}$ values ($1^-$, $2^-$, $2^+$, $3^-$, $3^+$, and $4^-$) that must be considered for each resonance, even at the lowest energies. Initially, the same technique as that in Ref. 10 was employed in an attempt to determine resonance spins. However, it soon became clear that the single ratio (hard singles/all coincidences) of Ref. 10 was not sufficient to separate all the $J^{\pi}$ values. Furthermore, it was clear that there was much more useful information in the pulse-height data, information that was not recorded in the experiment of Ref. 10.

Our task was made easier by the fact that we also had measured the total neutron cross section for $^{95}$Mo in a separate experiment, run on flight path 1 at ORELA using a $^6$Li-glass detector 80 m from the neutron production target. With the aid of these data, we were able to discern the parities of resonances having sufficiently large neutron widths. Examining pulse-height spectra for these resonances of known parity, it became clear that a ratio similar (soft coincidences/hard singles) to that of Ref. 10 as well as the ratio soft singles/hard singles could be used to separate resonances according to their spins, as shown in Fig. 2a. Further study revealed that resonances of different parity could be separated by using the ratio of a region of singles having intermediate pulse heights (intermediate singles) to the most energetic singles (extra hard singles), as shown in Fig. 2b. Other pulse-height ratios were found which also yielded good parity separation.

Although the resonance analysis is not yet complete, with this new technique we have already made a vast improvement in $J^{\pi}$ assignments over previous efforts. At present, we have been able to make 134 firm $J^{\pi}$ assignments out of 179 resonances observed below 5.4 keV. In contrast, in Ref. 10, only 10 firm $J^{\pi}$ assignments were made to 1.14 keV, one of which disagrees with our results. According to compilations[16], there are 13 firm $J^\pi$ assignments below 1.20 keV, all of which agree with our results. Finally, 32 firm $J^\pi$ assignments below 2.05 keV were made in a very recent experiment[17] employing the DANCE detector, two of which do not agree with our results. Finally, because part of the technique relies only on singles data, it can be applied to our previous data from ORELA. First tests on data for isotopes of Pt indicate that this new technique will be very useful for assigning both spins and parities.

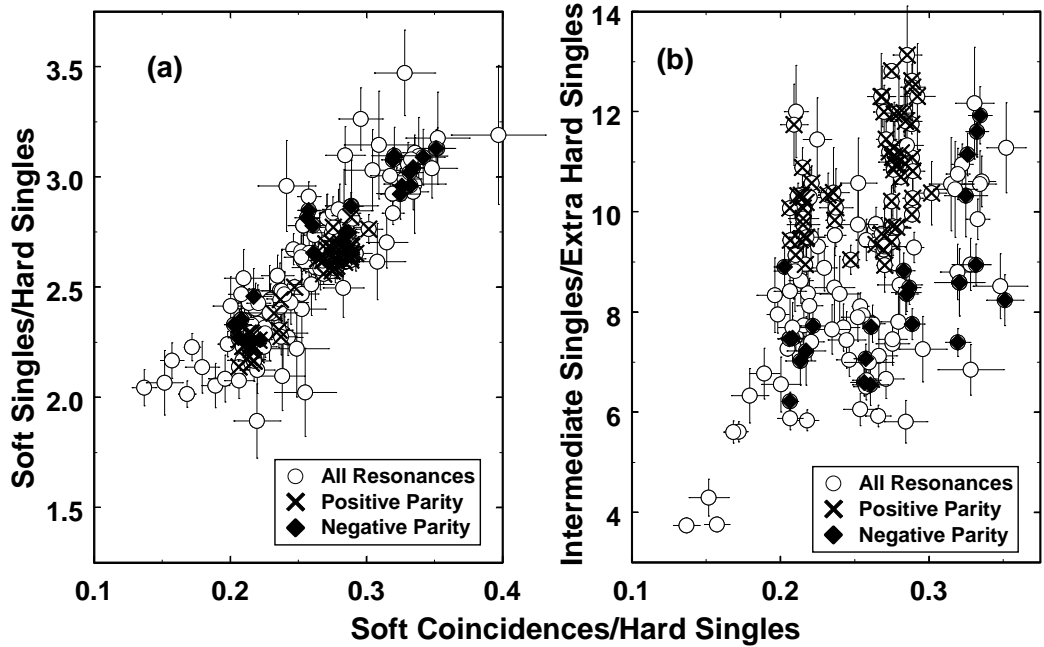

**FIGURE 2.** Ratios of counts in different pulse-height regions in singles and coincidence data taken at ORELA with a $^{95}$Mo sample. Circles represent all resonances below 7 keV. Filled diamonds and X's depict resonances identified as having negative or positive parity, respectively, by using only our ORELA transmission data. See text for details.

## NON-STATISTICAL EFFECTS IN $^{147}$Sm+$n$

As a result of our DANCE experiment, we were able to make firm $J^\pi$ assignments for virtually all resonances below 700 eV. Hence, we were able to obtain accurate average level spacings ($D_0$) and neutron strength functions ($10^4 S_0$) for both $J = 3$ and 4 resonances. The results were in good agreement with expectations of the statistical model and spin cutoff parameter. These $D_0$ and $S_0$ parameters uniquely determine the PT distribution with which the $\Gamma_n^0$ data should agree. However, applying the maximum-likelihood (ML) method of Ref. 1 to the $\Gamma_n^0$ data resulted in degrees-of-freedom values $\nu = 2.0\pm0.22$ and $1.5\pm0.22$ for $J = 3$ and 4, respectively, which are 4.5 and 2.3 standard deviations different from the expected value of $\nu = 1$ for a PT distribution.

Before proceeding further, we considered the fact that a non-statistical effect already had been reported[18] near $E_n = 350$ eV from analysis of $^{147}$Sm$(n,\alpha)$ data. With this in mind, we divided the reduced neutron width data into two groups from $E_n = 0$–350 eV and $E_n = 350$–700 eV. Also, because our analysis indicated that the average reduced neutron widths are equal for $J = 3$ and 4, we combined the data for the two spins to increase the statistical precision. The results are shown in Fig. 3, from which it can be seen that the distribution changes shape between the two energy regions. Applying the ML method of Ref. 1 to these data yielded $\nu = 1.02\pm0.22$ and $3.5\pm0.22$ for the 0–350- and 350–700-eV regions, respectively.

One problem with the method of Ref. 1 is that a rather artificial energy independent threshold is used to correct the data for missed resonances. Therefore, we applied two other[19,20], more realistic, methods to estimate the number of missed resonances. Both techniques indicated that very few resonances had been missed; fewer than five (of both spins) by 350 eV, and fewer than seven more in the 350–700-eV region. With these corrected number of resonances, we could apply the same ML technique that previously had been used[4] to demonstrate the validity of the PT distribution. However, in our case a very different conclusion was reached: The data change from being in agreement with PT ($\nu = 0.91\pm0.32$) in the lower-energy region to disagreeing with PT ($\nu = 3.19\pm0.83$) in the upper-energy region.

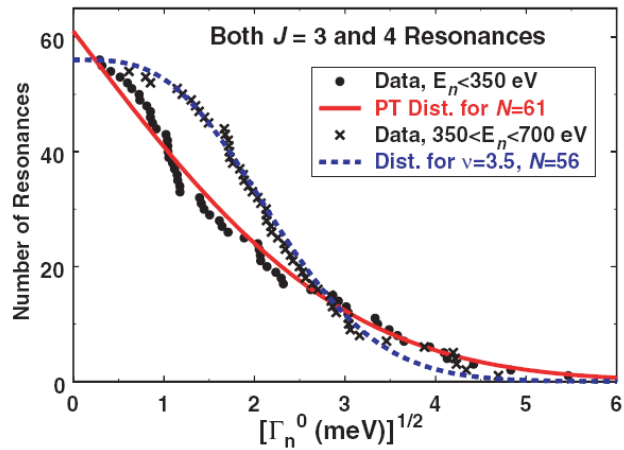

**FIGURE 3.** Distributions of reduced neutron widths for two different energy regions from our data taken with a $^{147}$Sm sample with the DANCE detector at LANSCE. Plotted are the number of resonances having a reduced neutron width greater than a given value versus the square root of that value. Resonance with $E_n<350$ eV and $350<E_n<700$ eV are shown as solid circles and X's, respectively. The sold and dashed curves are the expected PT and a $\nu=3.5$ distributions, respectively, after corrections for missed resonances. See text and Ref. 15 for details.

We also applied a second method, the Kolmogorov-Smirnov (KS) Test[21], to ascertain whether or not the data were in agreement with a PT distribution. Due to space limitations, we will not describe this method herein, but details can be found in Ref. 15. The basic result of applying the KS test to the data is the same as for the ML method: The data in the 350–700-eV region are inconsistent with a PT distribution to high statistical precision. For example, with the correction that there were seven missed resonances in this region, the KS test indicates that there is just a 0.003% chance that the data are consistent with a PT distribution. Even if the extremely conservative assumption is made that there were 17 resonances missed, the chance that the data are consistent with a PT distribution is still only 0.02%.

In Ref. 15, we discuss possible explanations for this strange non-statistical behavior of $^{147}$Sm+$n$ reduced neutron widths. In summary, we could find no reasonable explanation. When considered together with a similar effect reported[5-8] for $^{232}$Th+$n$, and the reported[18] non-statistical effect in the $^{147}$Sm(n,$\alpha$) reaction, both at nearly the same energy, it seems most likely that some unidentified nuclear structure effect, perhaps related to deformation, could be the cause. Similar deviations from a PT distribution have been reported[9] for five odd-A targets. In those cases however, there are very few spin assignments, so data for the two $s$-wave spins were combined in the analysis. It is possible that assumptions implicit to the way in which the data were combined are not valid. Therefore, new spin information, from new experiments applying the new techniques we have described herein, may be very useful for checking these reported effects and perhaps shedding more light on their origin.

## ACKNOWLEDGMENTS

This work was supported in part by the U.S. Department of Energy under Contract No. DE-AC05-00OR22725 with UT-Battelle, LLC. This work has benefited from the use of the LANSCE facility at Los Alamos National Laboratory, which was funded by the U.S. Department of Energy and operated by the University of California under Contract W-7405-ENG-36.